\documentclass[3p,times]{elsarticle}

\usepackage{ecrc}

\volume{00}

\firstpage{1}

\journalname{Nuclear Physics A}

\runauth{K. Dusling, T. Epelbaum, F. Gelis, R. Venugopalan}

\jid{nupha}

\usepackage{amssymb}

\usepackage[figuresright]{rotating}
\usepackage{etex}
\usepackage{psfrag}
\usepackage{graphics}
\usepackage{pstricks}
\usepackage[english]{babel}
\usepackage{hyperref}
\usepackage{tikz}
\usepackage{colortbl}
\usepackage{yfonts}
\usepackage{bbm}
\usepackage{amsmath}
\usepackage{multicol}

\def\colora{}
\def\colorb{}
\def\colorc{}
\def\colord{}

\def\empile#1\over#2{\mathrel{\mathop{\kern 0pt#1}\limits_{#2}}}
\newcommand{\slvarepsilon}{\raise.15ex\hbox{$/$}\kern-.53em\hbox{$\varepsilon$}}
\newcommand{\slL}{\raise.15ex\hbox{$/$}\kern-.53em\hbox{$L$}}
\newcommand{\slP}{\raise.15ex\hbox{$/$}\kern-.53em\hbox{$P$}}
\newcommand{\slD}{\raise.15ex\hbox{$/$}\kern-.53em\hbox{$D$}}
\newcommand{\slp}{\raise.1ex\hbox{$/$}\kern-.63em\hbox{$p$}}
\newcommand{\slq}{\raise.1ex\hbox{$/$}\kern-.53em\hbox{$q$}}
\newcommand{\slv}{\raise.1ex\hbox{$/$}\kern-.63em\hbox{$v$}}
\newcommand{\slR}{\raise.15ex\hbox{$/$}\kern-.53em\hbox{$R$}}
\newcommand{\slQ}{\raise.15ex\hbox{$/$}\kern-.53em\hbox{$Q$}}
\newcommand{\slK}{\raise.15ex\hbox{$/$}\kern-.53em\hbox{$K$}}
\newcommand{\slk}{\raise.15ex\hbox{$/$}\kern-.53em\hbox{$k$}}
\newcommand{\slSigma}{\raise.15ex\hbox{$/$}\kern-.53em\hbox{$\Sigma$}}
\newcommand{\slcalP}{\raise.15ex\hbox{$/$}\kern-.63em\hbox{$\cal P$}}
\newcommand{\slcalA}{\raise.15ex\hbox{$/$}\kern-.73em\hbox{$\cal A$}}
\newcommand{\slA}{\raise.15ex\hbox{$/$}\kern-.73em\hbox{$A$}}
\newcommand{\slbfA}{\raise.15ex\hbox{$/$}\kern-.73em\hbox{${\imb A}$}}
\newcommand{\slpartial}{\raise.15ex\hbox{$/$}\kern-.53em\hbox{$\partial$}}
\newcommand{\sla}{\raise.15ex\hbox{$/$}\kern-.53em\hbox{$a$}}
\newcommand{\slb}{\raise.15ex\hbox{$/$}\kern-.53em\hbox{$b$}}
\newcommand{\slc}{\raise.15ex\hbox{$/$}\kern-.53em\hbox{$c$}}
\newcommand{\slC}{\raise.15ex\hbox{$/$}\kern-.63em\hbox{$C$}}

\def\p{{\boldsymbol p}}

\def\k{{\boldsymbol k}}

\def\u{{\boldsymbol u}}
\def\v{{\boldsymbol v}}

\def\bs{\boldsymbol}

\begin{document}

\begin{frontmatter}



\dochead{}

\title{{\bf Initial state and thermalization}}


\author[ncsu]{K. Dusling}
\author[saclay]{T. Epelbaum}
\author[saclay]{F. Gelis}
\author[bnl]{R. Venugopalan}

\address[ncsu]{Physics Department, North Carolina State University,
    Raleigh, NC 27695, USA}
\address[saclay]{Institut de Physique Th\'eorique,
   CEA, 91191 Gif-sur-Yvette Cedex, France}
\address[bnl]{Physics Department,
   Brookhaven National Laboratory, Upton, NY-11973, USA}

\begin{abstract}
  We report recent results on the role of instabilities
  in the isotropization and thermalization of a longitudinally
  expanding system of quantum fields.
\end{abstract}


\end{frontmatter}

\section{Introduction}
\label{sec:intro}
The evolution of the quark-gluon matter produced in ultrarelativistic
heavy ion collisions is apparently well described by hydrodynamics
with a very small viscosity. Given its derivation from an expansion of
the energy-momentum tensor in gradients of the velocity field, one
expects hydrodynamics to be applicable only when the transverse and
longitudinal pressures are sufficiently close, and if the system is
close enough to local thermal equilibrium.

An outstanding problem in the application of Quantum ChromoDynamics to
the description of the early stages of these collisions is to justify
that these conditions are indeed met at a short enough time to apply
hydrodynamics. The main QCD tool for performing these studies is the
Color Glass Condensate (CGC) framework~\cite{CGC}, an effective theory
designed to describe the highly occupied gluon states encountered in
the wave functions of high energy nuclei, and their interactions in a
collision. In the CGC, the fast partons ($k^+>\Lambda^+$ for a nucleus
moving in the $+z$ direction) are described as static (thanks to time
dilation of all the internal dynamics in a high energy nucleus) color
sources $J^\mu\sim \delta^{\mu +}\rho$ on the light-cone, with a
probability $W[\rho]$. The gluons below the cutoff ($k^+<\Lambda^+$)
are described as ordinary gauge fields, coupled eikonally to the fast
sources.
When this effective description is applied to a collision, the current
$J^\mu$ is the sum of two terms, corresponding respectively to each of
the colliding nuclei.

\section{Initial state factorization}
\label{sec:ini}
An important issue in applications of the CGC to the calculation of
observable quantities is the dependence on the cutoffs that are
introduced to separate the fast and slow partons. This cutoff is not a
physical quantity and should cancel in observables. However, since it
is the upper bound in loop integrals, higher order corrections usually
produce logarithms of the cutoff. It is well known that in
electron--nucleus collisions, these logarithms can be absorbed by
letting the distribution $W[\rho]$ depend on the scale $\Lambda^+$
according to the JIMWLK equation \cite{JIMWLK}. For the CGC to provide
a consistent description of nucleus--nucleus collisions, one needs to
prove that the same distributions $W[\rho]$'s are sufficient to absorb
all the logarithms that arise in these collisions, and that this works
for a sufficiently large class of observables.

The first step in this proof is to show that at Leading Order (LO) in
the strong coupling $\alpha_s$, all the inclusive observables
(i.e. those that do not restrict the content of the final state, such
as particle spectra, or the expectation value of some local operators
in the final state) can be expressed in terms of the retarded solution
of the classical Yang-Mills equations that vanishes in the remote past
($x^0\to -\infty$) \cite{GelisV2}. For instance, in terms of this
classical field ${\cal A}^\mu$, the gluon spectra at LO read
\begin{equation}
\left.
\frac{dN_1}{d^3{\colorb\vec\p}}\right|_{_{\rm LO}}
\sim
\int d^4x d^4y\; e^{ip\cdot(x-y)}\;\square_x \square_y \;
{\colord{\cal A}(x)\,{\cal A}(y)}
\; ,\quad
\left.
\frac{dN_n}{d^3{\colorb\vec\p_1}\cdots d^3{\colorb\vec\p_n}}\right|_{_{\rm LO}}
=
\left.
\frac{dN_1}{d^3{\colorb\vec\p_1}}\right|_{_{\rm LO}}\cdots
\left.
\frac{dN_1}{d^3{\colorb\vec\p_n}}\right|_{_{\rm LO}}\; .
\end{equation}

The next step is to compute the Next to Leading Order (NLO) correction
to these observables. In \cite{NLO}, we have shown that
for an inclusive observable ${\cal O}$, the LO and NLO contributions
are formally related by
\begin{equation}
{\colorb{\cal O}}_{_{\rm NLO}}
=
\Bigg[
 \frac{1}{2}
\int\limits_{_{\u,\v}}
\int_\k
\big[{\colorb {\bs a}_\k}\,{\mathbbm T}\big]_\u 
\big[{\colorb {\bs a}_\k^*}\,{\mathbbm T}\big]_\v
+\int\limits_{_{\u}}
\big[{\colorb{\bs\alpha}}\,{\mathbbm T}\big]_\u
\Bigg]\;
{\colorb{\cal O}}_{_{\rm LO}}\; .
\label{eq:LO-NLO}
\end{equation}
In this formula, ${\mathbbm T}$ is the operator that generates shifts
of the classical field at some initial time (and the integrals over
$\u,\v$ are integrals over the spatial coordinates at this time). The
quantities $a_\k,\alpha$ are small perturbations to the classical
field, evaluated at the same time. The formula is true for any initial
time, but some choices simplify the problem by making $a_\k,\alpha$
calculable analytically. Then, the main result in proving the
factorization of the logarithms of the cutoffs is
\begin{equation}
    \frac{1}{2}
\int\limits_{_{\u,\v}}
\int_\k
\big[{\colorb {\bs a}_\k}\,{\mathbbm T}\big]_\u 
\big[{\colorb {\bs a}_\k^*}\,{\mathbbm T}\big]_\v
+\int\limits_{_{\u}}
\big[{\colorb{\bs\alpha}}\,{\mathbbm T}\big]_\u
=
    \log\left(\Lambda^+\right)\,{\colorb{\cal H}_1}
    +
    \log\left(\Lambda^-\right)\,{\colorb{\cal H}_2}
    +
    \mbox{terms w/o logs}\; ,
\end{equation}
where ${\cal H}_{1,2}$ are the JIMWLK Hamiltonians of the two
nuclei. This formula shows that the logarithms of the two cutoffs do
not mix, since they are multiplied by operators that act on a single
nucleus. From this formula, it is easy to show that one can absorb all
the leading logarithms in two distributions $W_{1,2}$ that evolve with
their respective JIMWLK equation,
\begin{equation}
\left<{\colorb{\cal O}}\right>_{_{\rm Leading\ Log}}
=
\int 
\big[D{\colora\rho_{_1}}\,D{\colorb\rho_{_2}}\big]
\;
{\colora W_1\big[\rho_{_1}\big]}\;
{\colorb W_2\big[\rho_{_2}\big]}
\;
{{\colorb{\cal O}}_{_{\rm LO}}[\rho_1,\rho_2]}\; .
\end{equation}
It is the universality of the distributions $W$ (i.e. the fact that
the same distributions enter in different types of reactions involving
the same nuclei) that gives its predictive power to the CGC.

\section{Thermalization and isotropization}
\label{sec:iso}

\subsection{Resummation of the unstable modes}
\begin{figure}[htbp]
\begin{center}
\resizebox*{!}{3cm}{\includegraphics{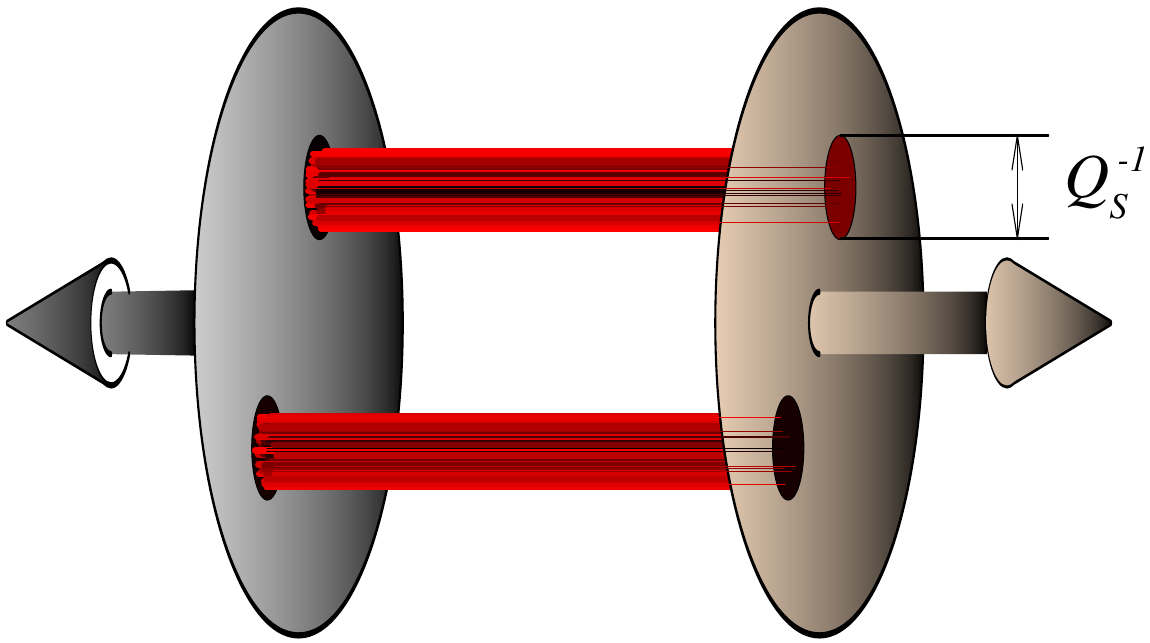}}\hfil
\resizebox*{!}{3cm}{\rotatebox{0}{\includegraphics{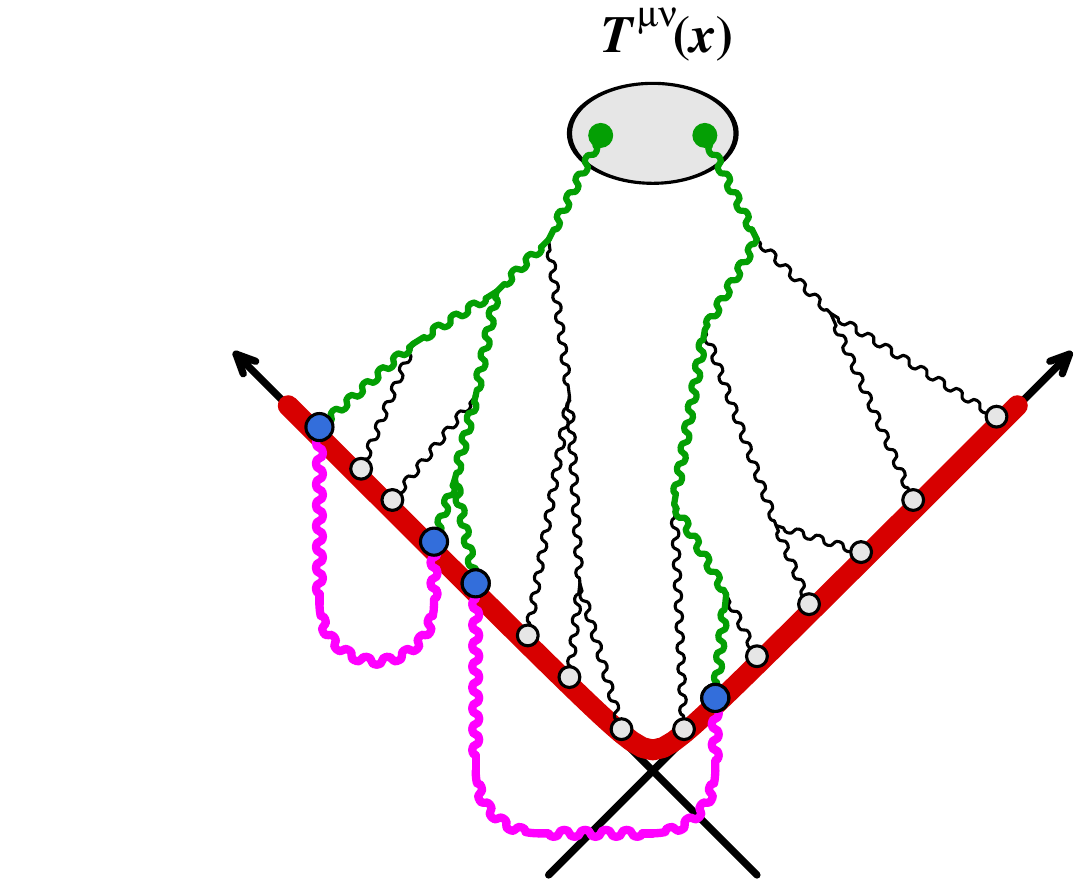}}}\hfil
\resizebox*{!}{3cm}{\rotatebox{0}{\includegraphics{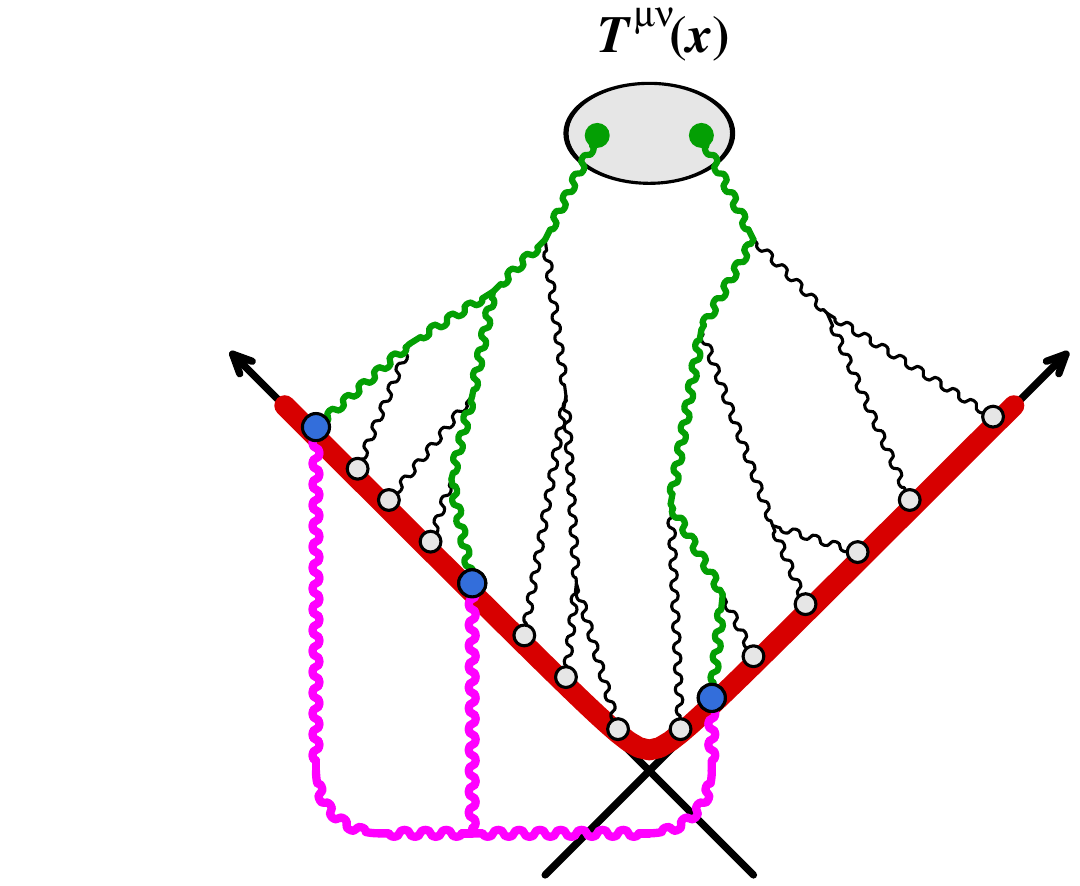}}}
\end{center}
\caption{\label{fig:tubes}Left: color flux tubes. Right: examples of
  2-loop corrections.}
\end{figure}

At LO, the classical color field produced in a nucleus-nucleus
collisions corresponds to longitudinal chromo-electric and
chromo-magnetic fields (figure \ref{fig:tubes} - left) that form {\sl
  flux tubes} whose transverse size is of the order of the inverse
saturation momentum. The energy momentum tensor of this field
configuration has a negative longitudinal pressure, and is thus quite
far from what can be reasonably described by hydrodynamics.

Moreover, it has been known for some time that the classical solutions
of Yang-Mills equations have instabilities: small rapidity dependent
perturbations to the boost invariant classical field encountered at LO
grow exponentially in time, and eventually become as large as the
classical field itself \cite{insta}. Such
instabilities occur in loop corrections and lead to secular
divergences in observables. These unstable contributions can be
tracked by an improvement of the power counting, where one assigns an
exponential of time to each operator ${\mathbbm T}$. Then one
sees that the terms with the strongest divergences (one
exponential factor per power of the coupling $g$ -- see figure
\ref{fig:tubes}, right, for examples of 2-loop graphs that differ in
the strength of their secular divergences) are obtained by
exponentiating the operator in eq.~(\ref{eq:LO-NLO}).
This exponentiation performs a resummation of a selected subset of all
the higher loop corrections. When applied to the energy-momentum
tensor, this resummation reads
\begin{eqnarray*}
{\colorb T^{\mu\nu}_{_{\rm resummed}}}
&=&
{\colora\exp}\Bigg[
\frac{1}{2}\int\limits_{_{\u,\v}}
\underbrace{{\colorc\int_\k}
[{\colorc{\bs a}_\k}{\mathbbm T}]_\u
[{\colorc{\bs a}_\k^*}{\mathbbm T}]_\v}_{{\colorc{\cal G}(\u,\v)}}
+
\int\limits_{_{\u}}
[{\colorb{\bs\alpha}}{\mathbbm T}]_\u
\Bigg]\;
 {T^{\mu\nu}_{_{\rm LO}}}[{\cal A}_{\rm init}]
\nonumber\\
&=&
\int[D{\colord\chi}]\,
\exp\Bigg[-\frac{1}{2}
\int\limits_{_{\u,\v}}
{\colord\chi(\u)}{\colorc{\cal G}^{-1}(\u,\v)}{\colord\chi(\v)}\Bigg]\;
T^{\mu\nu}_{_{\rm LO}}[{\cal A}_{\rm init}+{\colord\chi}+{\colorb\alpha}]\; .
\end{eqnarray*}
The second line is an identity, based on the fact that the operator
${\mathbbm T}$ is the generator of the shifts of the initial classical
field ${\cal A}_{\rm init}$. It shows that this resummation amounts to
a Gaussian average over fluctuating initial conditions, centered on
the classical one.

\subsection{Non expanding system}
In order to test the effect of this resummation in a simpler setting,
we have studied a scalar field theory with a $\phi^4$
interaction. This theory has also unstable classical solutions, due to
parametric resonance. Moreover, in four space-time dimensions, this
theory is scale invariant at the classical level, as is Yang-Mills
theory.

\begin{figure}[htbp]
\begin{center}
\resizebox*{!}{4cm}{\rotatebox{-90}{\includegraphics{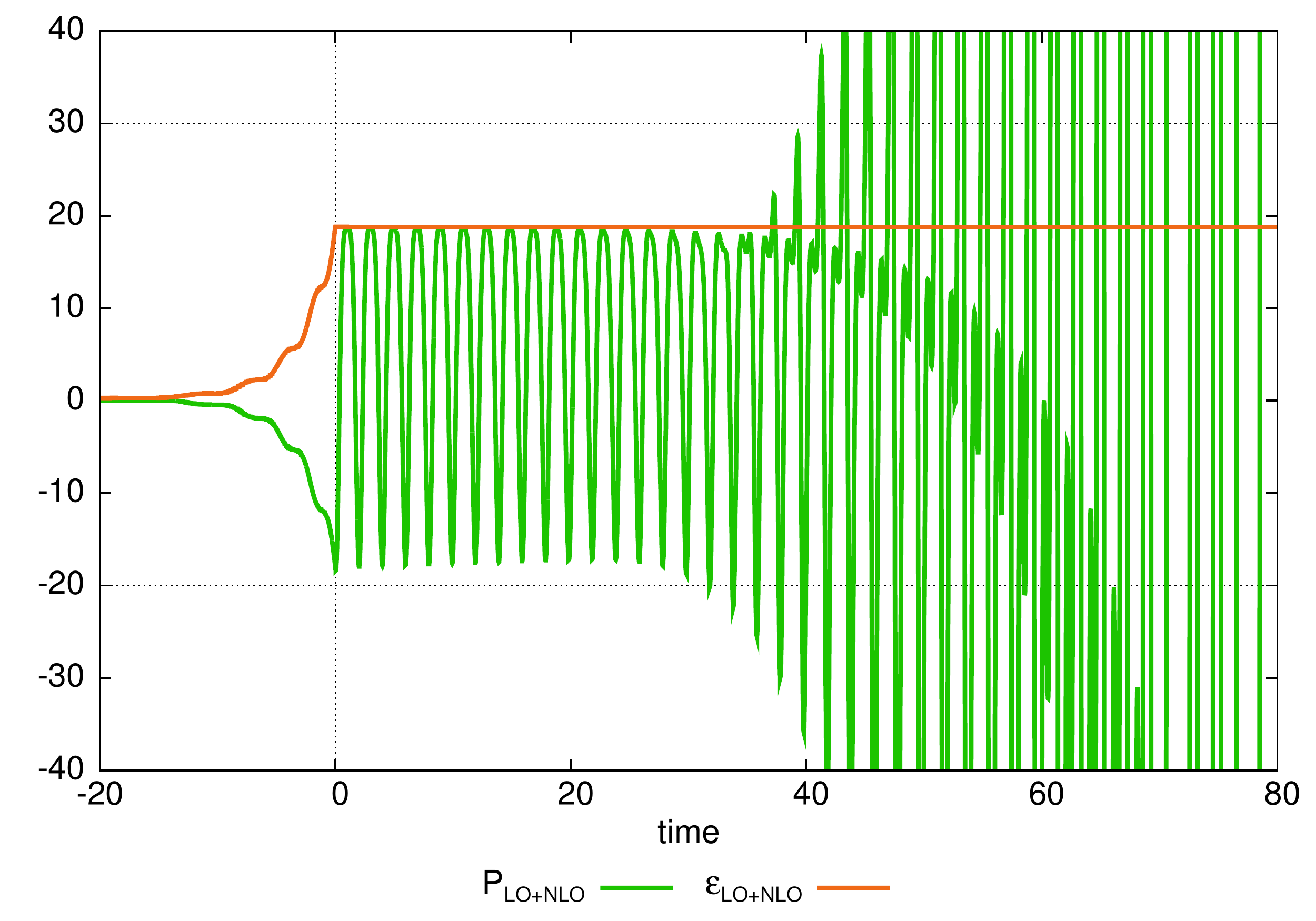}}}\hfil
\resizebox*{!}{4cm}{\rotatebox{-90}{\includegraphics{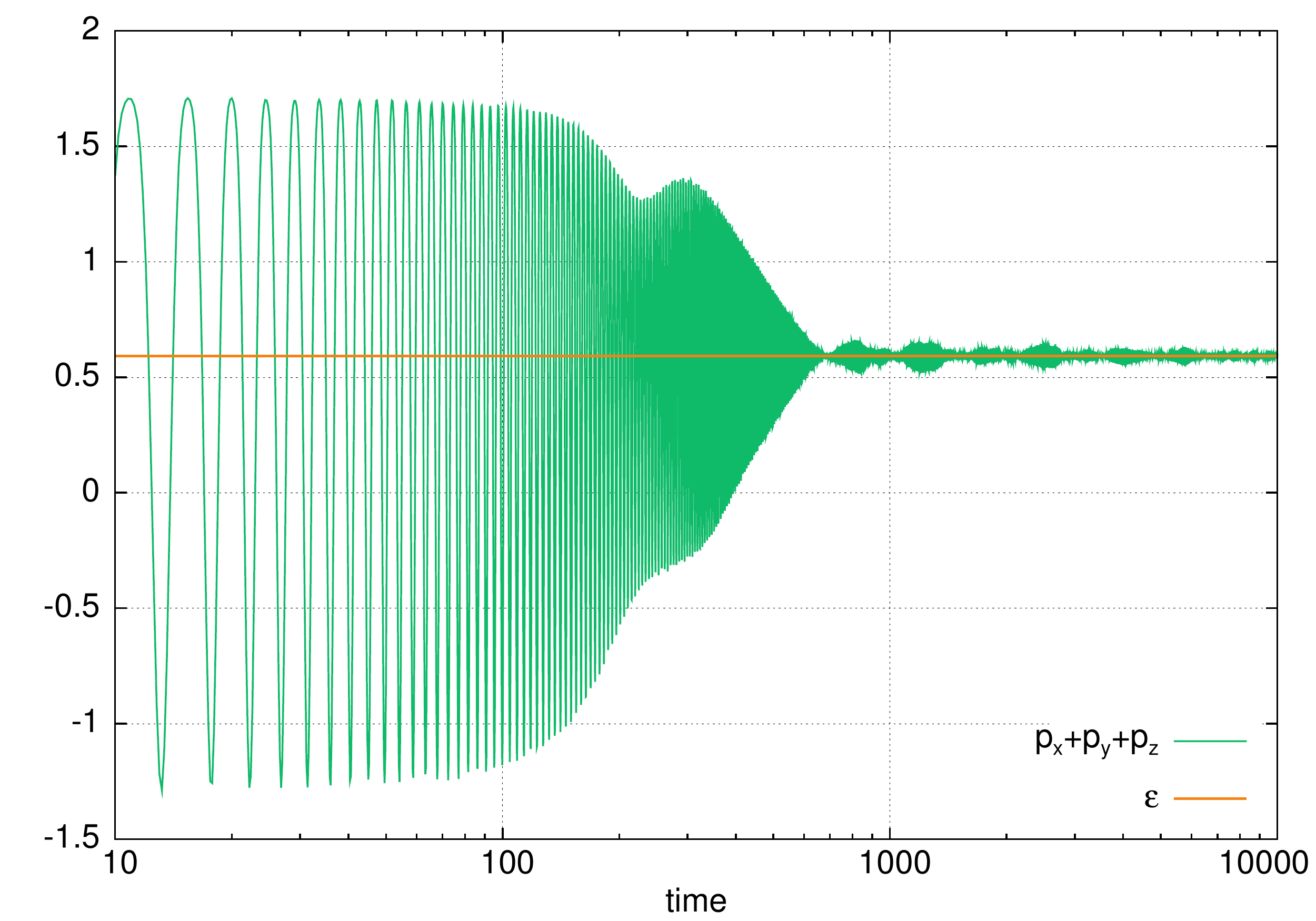}}}
\end{center}
\caption{\label{fig:resum}Left: energy density and pressure at LO and
  NLO. Right: resummed energy density and pressure.}
\end{figure}

We have first considered this theory in the case of a fixed volume
\cite{fixed}. First of all, one can check explicitly the
presence of secular divergences in fixed loop order computations of
the energy-momentum tensor (see figure \ref{fig:resum}, left). On the
right plot of this figure, we also see that the behavior is quite
different after the resummation: not only has the secular divergence
been cured, but one also sees that the pressure relaxes to the
equilibrium value $P=\epsilon/3$.

\begin{figure}[htbp]
\begin{center}
\resizebox*{!}{4cm}{\rotatebox{-90}{\includegraphics{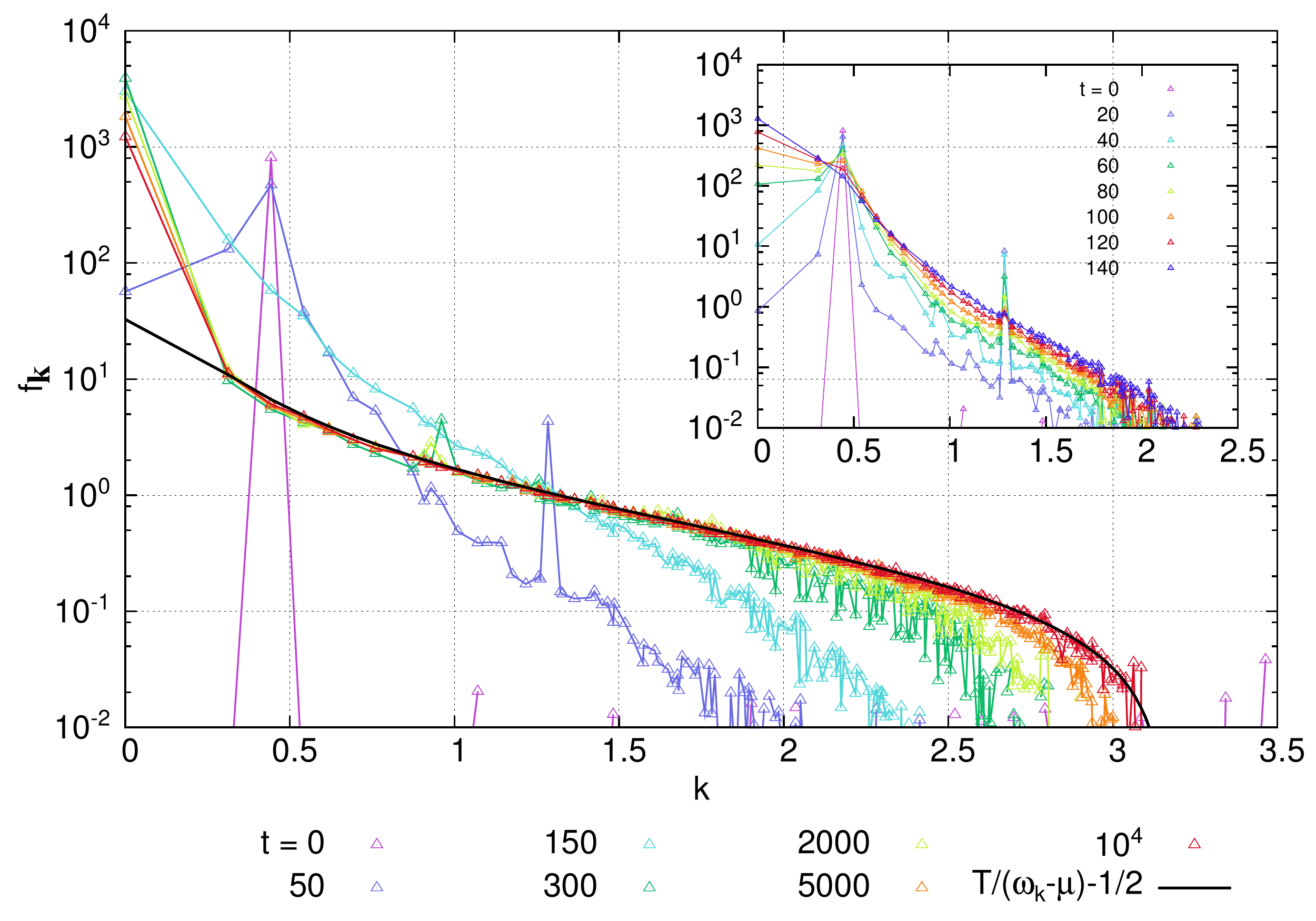}}}\hfil
\resizebox*{!}{4cm}{\rotatebox{-90}{\includegraphics{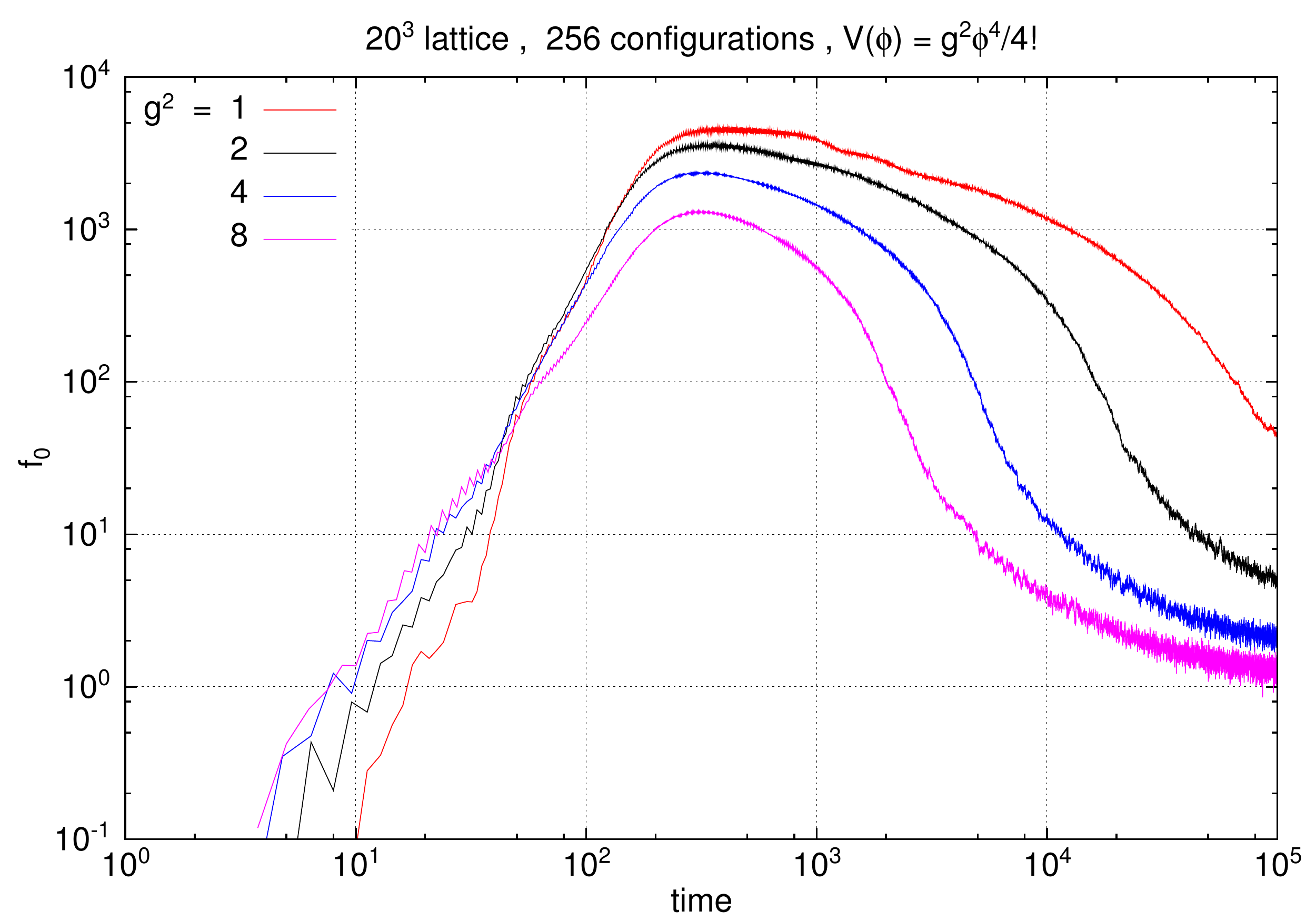}}}
\end{center}
\caption{\label{fig:fk}Left: evolution of the particle
  distribution. Right: time evolution of the occupation number at
  ${\boldsymbol k}=0$.}
\end{figure}

In this set-up, one can also compute the occupation number at various
times in the evolution (figure \ref{fig:fk}, left).
In this computation, the initial classical field was set so that the
occupation number is large in a single momentum mode, and zero
elsewhere. Rapidly, a population of particles builds up in the
neighboring modes. At late times, the distribution has evolved into a
classical equilibrium. Interestingly, the best fit requires a positive
chemical potential. This indicates that chemical equilibration has not
yet happened at these times, suggesting that the number changing
processes are considerably slower than the elastic ones.  Moreover,
this chemical potential is almost equal to the mass of the quasi-particles,
and sees a marked excess of particles in the zero mode, both
indicative of Bose-Einstein condensation (BEC). By following the occupation
of the zero mode over longer times (figure \ref{fig:fk}, right), one
sees that indeed this BEC is only transient and disappears eventually
when inelastic processes equilibrate the particle number. Whether a
similar phenomenon can occur for gauge fields is still an open issue
\cite{BEC}.

\subsection{Longitudinal expansion}
\begin{figure}[htbp]
\begin{center}
\resizebox*{!}{4cm}{\rotatebox{0}{\includegraphics{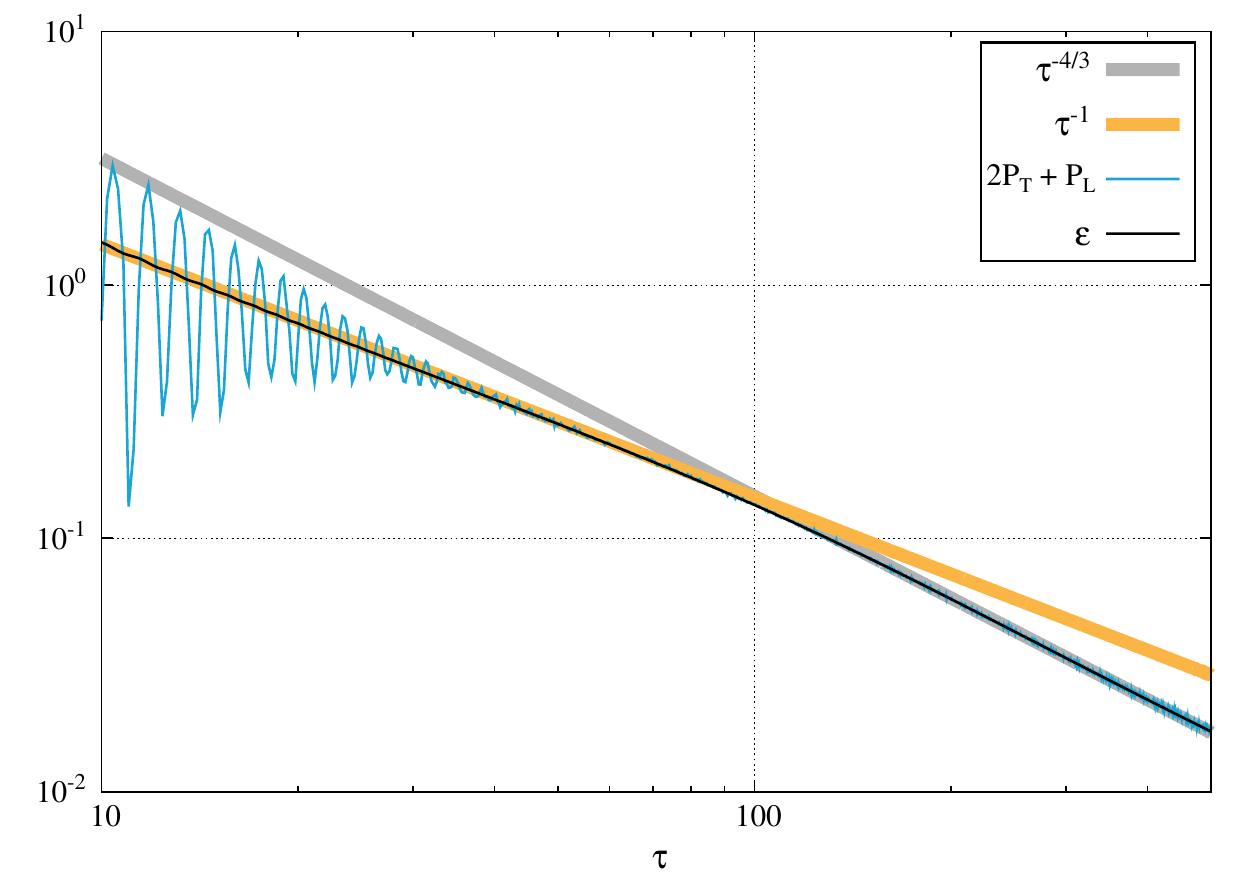}}}\hfil
\resizebox*{!}{4cm}{\rotatebox{0}{\includegraphics{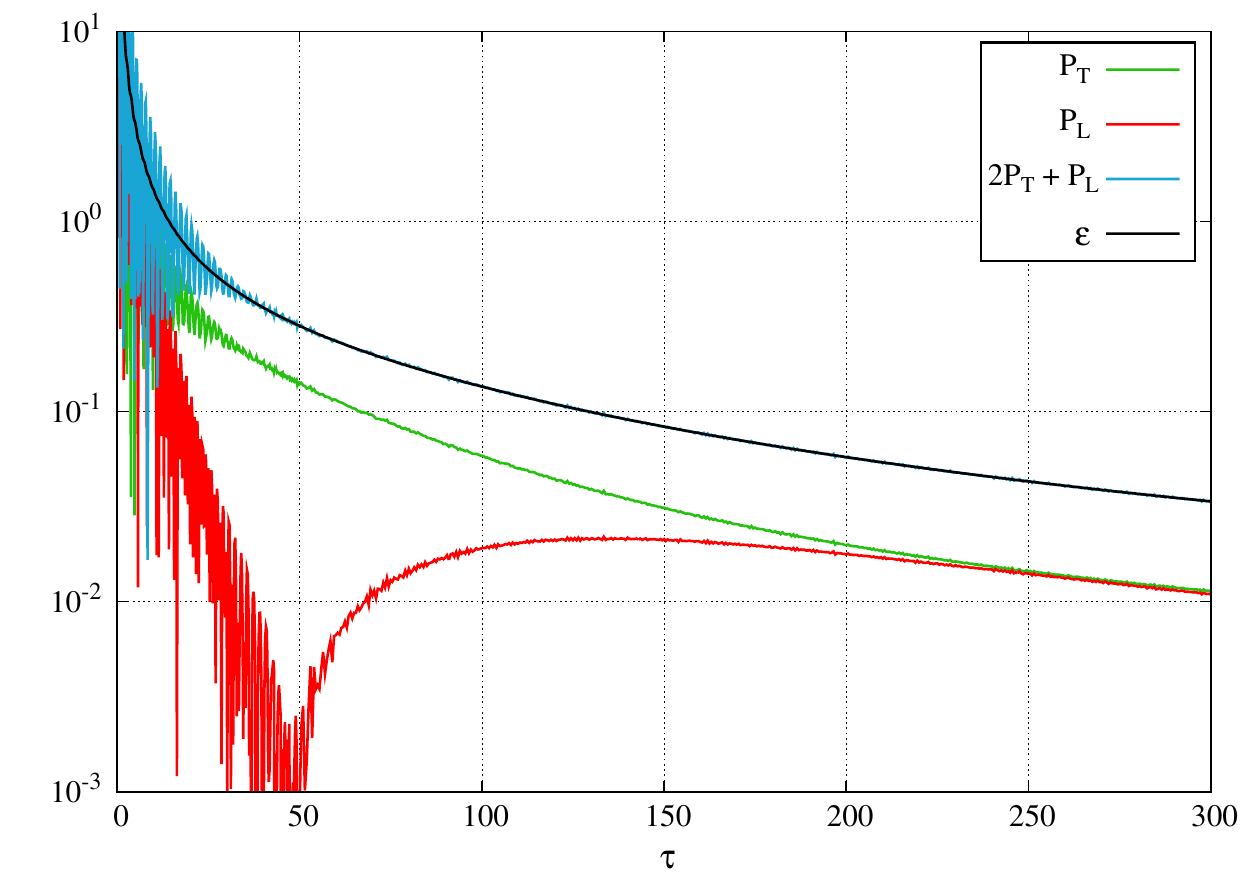}}}
\end{center}
\caption{\label{fig:iso}Left: relaxation of the trace of
  $T^{\mu\nu}$. Right: isotropization of the pressures.}
\end{figure}
The previous scalar model can be studied also for a longitudinally
expanding system, where the classical background field is boost
invariant \cite{DusliEGV2}. In the left panel of figure
\ref{fig:iso}, one first sees that the trace of the pressure tensor
relaxes towards the energy density. Moreover, the time dependence of
the energy density changes from $\tau^{-1}$ to $\tau^{-4/3}$,
indicating a qualitative change from a very small longitudinal
pressure to a longitudinal pressure equal to $\epsilon/3$. This is
confirmed in the right panel of figure \ref{fig:iso}: after a very
rapid initial decrease, due to the redshifting of the longitudinal
momenta, the longitudinal pressure starts growing again due to the
instabilities, to eventually become very close to the transverse
pressure. The spectrum of fluctuations appropriate for gauge fields
over a boost invariant background has been worked out in
\cite{DusliGV1}, paving the way for the same simulation in Yang-Mills
theory. \vglue 2mm

\noindent{\bf Acknowledgements\ \ } F.G. and T.E. are
supported by the Agence Nationale de la Recherche project
\#~11-BS04-015-01. R.V.'s work is supported by the US Department of
Energy under DOE Contract No.DE-AC02-98CH10886 and by an LDRD grant
from Brookhaven Science Associates.


\end{document}